\documentclass[fleqn,12pt,twoside]{article}
\usepackage{espcrc1}
\usepackage{graphicx}
\newcommand{\AmS}{{\protect\the\textfont2
  A\kern-.1667em\lower.5ex\hbox{M}\kern-.125emS}}
\newcommand{\eq}{{\,=\,}}
\newcommand{\e}{\varepsilon}

\title{The quark-gluon plasma at RHIC\thanks{This work was supported by 
       the U.S. Department of Energy under contract DE-FG02-01ER41190.}
       }

\author{Ulrich Heinz\\[1ex]
        Department of Physics, The Ohio State University, Columbus, 
        OH 43210, USA}

\begin{document}

\maketitle

\begin{abstract}
I present a theory-guided review of RHIC data, arguing that they provide
strong evidence for formation of a thermalized quark-gluon plasma at RHIC.
Strong radial flow reflects high thermal pressure in the reaction zone.
Large elliptic flow proves that the pressure builds up quickly and the
system thermalizes on a very short time scale of ${<\,}1$\,fm/$c$. The 
observed hadron yields are consistent with statistical hadron formation 
from a quark-gluon plasma, followed by immediate chemical decoupling due 
to strong radial expansion. The observed suppression of jets appears to
confirm the predicted large energy loss suffered by hard partons moving
through a quark-gluon plasma; more work is required to quantitatively
understand this effect. Source size measurements using two-particle
correlations do not seem to fit into this picture; the origin of this 
discrepancy (``HBT puzzle'') is presently not understood.
\end{abstract}



\section{THE LITTLE BANG: QGP AT RHIC}
\label{sec1}

\vspace*{-0.1cm}

A quark-gluon plasma (QGP) is a thermalized system of deconfined quarks,
antiquarks, and gluons. As such it has thermodynamic pressure $P=c_s^2 \e$
where $\e$ is the energy density and $c_s$ is the sound velocity. 
Perturbative QCD gives $c_s^2\eq1/3$ at leading order, and lattice QCD 
confirms \cite{Karsch_QM01} that for $T{\,\geq\,}2 T_c$ (where 
$T_c{\,\approx\,}170$\,MeV is the critical temperature for color deconfinement 
in QCD \cite{Karsch_QM01}) about 80-85\% of this value is reached. In heavy
ion collisions the reaction zone is surrounded by vacuum with zero pressure;
thus, if the collision fireball contains a QGP, the pressure gradient near 
the surface will lead to {\em collective expansion (``flow'')}, in particular
transverse to the beam in which direction the nuclear matter was initially 
at rest. Therefore, collective transverse flow is an unavoidable 
consequence of QGP formation, and the data must show it if a QGP has 
been created.

The converse is not necessarily true unless the observed flow is so strong
that only a QGP could have generated sufficient pressure over a sufficiently
long time to create it. In order to assess whether the latter is the case 
one exploits three facts: 

(i) Due to incomplete stopping of the two 
nuclei at high collision energies, the reaction zone expands 
quickly in the beam direction, thereby rapidly cooling and diluting the 
matter inside; hence the pressure has only a limited time interval 
available to generate flow in the transverse directions. This time is 
the shorter the lower the initial energy density; one can therefore 
establish an upper limit for the amount of ``radial'' transverse flow 
that can be generated if the initial energy density (at the time of 
approximate thermalization) was never significantly above the critical 
value for color deconfinement (about 1 GeV/fm$^3$).

(ii) By energy conservation, the initial energy density is related to
the final transversally emitted energy per unit rapidity, $dE_T/dy$, via
geometric factors and the time of thermalization \cite{Bj83}: the shorter 
the thermalization time, the higher the initial energy density at fixed 
(measured) final $dE_T/dy$. The Bjorken formula \cite{Bj83}
$\e(\tau_{\rm therm})\eq\frac{1}{\tau_{\rm therm}} 
 \frac{1}{\pi \langle R^2\rangle} \frac{dE_T}{dy}$ is actually an 
underestimate of the initial energy density since it neglects longitudinal
work done by the pressure during the expansion \cite{HK85}. The measured
values for $dE_T/dy$ at RHIC \cite{E_T} are consistent with subcritical
initial energy densities only if one assumes that thermalization takes
at least $\tau_{\rm therm}{\,\geq\,}5$\,fm/$c$.

(iii) On the other hand, this thermalization time scale can be constrained 
by the measured {\em elliptic flow} in noncentral collisions 
\cite{Sorge:1997pc}. In such collisions, the nuclear reaction zone is 
initially deformed in coordinate space: a cut transverse to the beam 
direction looks like an almond whose longer side points perpendicular 
to the reaction plane. For a given collision centrality, extracted e.g. 
from the 
measured total multiplicity $dN/dy$, this deformation can be calculated
from the overlap geometry \cite{KHHET}. Thermalization then leads to 
pressure gradients which are {\em anisotropic} and larger in the short 
direction of the almond, causing a faster growth of the transverse flow 
into the reaction plane than perpendicular to it. This mechanism leads 
to an anisotropy in the final transverse momentum distribution, making 
it flatter in the direction of the impact parameter vector than 
perpendicular to it \cite{O92}. The {\em elliptic flow coefficient} 
$v_2$, defined as the second harmonic coefficient in an azimuthal 
Fourier decomposition of the measured transverse momentum spectrum, 
quantifies this anisotropy. Without reinteractions among the produced 
particles, no such momentum anisotropy would arise; rescattering in the 
hot fireball matter is required to transfer the initial spatial 
anisotropy into momentum space. Microscopic calculations 
\cite{Molnar:2001ux} show that, for a given initial spatial deformation, 
the generated amount of $v_2$ is a monotonic function of the mean 
free path (or the product of density and scattering cross section) in 
the fireball; the maximal value is reached in the hydrodynamic limit of 
zero mean free path \cite{Heinz:2001xi}. If thermalization is delayed, 
the initial spatial deformation disappears spontaneously by transverse 
free-streaming \cite{Kolb:2000sd}, thereby reducing the maximal 
possible momentum anisotropy. Thus, the measured elliptic flow at given 
collision centrality provides immediately an upper limit for the 
thermalization time scale, with larger $v_2$ values corresponding to 
smaller values for $\tau_{\rm therm}$. In the hydrodynamic limit 
the extracted $\tau_{\rm therm}$ depends parametrically on the stiffness
of the equation of state (velocity of sound $c_s^2$), with smaller sound
velocities resulting in shorter thermalization times at fixed $v_2$
\cite{Kolb:2000sd}. The weakest upper limit for $\tau_{\rm therm}$ is 
thus extracted from the $v_2$ data by taking the largest reasonable 
value for the speed of sound, $c_s^2\eq\frac{1}{3}$, corresponding to an
ideal gas of massles quarks and gluons. 

Experimentally, the strength of the radial flow can be extracted from the
transverse mass ($m_\perp\eq\sqrt{m^2+p_\perp^2}$) spectra of a variety
of different mass hadrons by fitting them to a common flow spectrum 
\cite{Lee:1990sk}. At high $m_\perp{\,\gg\,}m$, the transverse flow flattens 
the thermal spectra by a simple common blueshift factor, and all spectra 
approach the same asymptotic slope from which the decoupling temperature
and flow velocity cannot be separated. At low $m_\perp{\,<\,}2m$, however, 
the flow induces an even stronger flattening which increases with the rest 
mass of the particles. Heavier hadrons thus develop a visible shoulder 
at small $m_\perp$, and this feature can be used to determine the average
transverse flow velocity uniquely. An excellent systematic study of the
shapes of the $m_\perp$ spectra of $\pi^\pm,K^\pm,\phi,p,\bar p,\Lambda,
 \bar\Lambda, \Xi, \bar \Xi, \Omega, \bar\Omega$, and $d$ from Pb+Pb 
collisions at beam energies of 40, 80 and 160 $A$\,GeV at the SPS was 
recently presented by M. van Leeuwen at the {\em Quark Matter 2002} 
conference \cite{Afanasiev:2002fk}, showing perfect consistency 
of all these spectra with a two-parameter flow fit \cite{Lee:1990sk}
with kinetic freeze-out temperature $T_{\rm f}\eq120-130$\,MeV and 
average transverse flow velocity $\langle\beta_\perp\rangle\eq0.4-0.5$ 
($T_{\rm f}$ and $\langle\beta_\perp\rangle$ are anticorrelated in these 
fits). Even the $\Omega$ and $\bar\Omega$ show a clear flow shoulder and
are consistent with these low freeze-out temperature and large flow values.
This contradicts earlier conclusions drawn from simple exponential fits 
to $\Omega$ spectra measured at higher $m_\perp$ values by the WA97 
Collaboration \cite{Antinori:2000sb} which did not show clear evidence 
for the flow shoulder and which led to the suggestion 
\cite{vanHecke:1998yu} that $\Omega$ and $\bar\Omega$ are too heavy and 
too weakly coupled to the expanding pion fluid to pick up much of the 
transverse flow created during the late hadronic stages of the collision, 
allowing them to decouple earlier, i.e. at higher temperature and smaller 
flow. The new data \cite{Afanasiev:2002fk} show that the $\Omega$ and 
$\bar\Omega$ fully participate in the flow and essentially decouple 
together with the rest of the hadron fluid. They also demonstrate the 
fallacies connected with characterising the spectra by a single slope 
parameter instead of comparing their entire non-exponential shape with 
a proper flow parametrization.

The hadron data at RHIC are already almost as rich as at the SPS, but
not yet all published in final form. Available flow analyses of pion, 
kaon and (anti)proton spectra \cite{XuQM01,Burward-Hoy:2002te,Peitzmann:2002nt}
give a similar range of freeze-out temperatures and average flow velocities
as at the SPS, but in a different combination (either $T_{\rm f}$ or
$\langle\beta_\perp\rangle$ is about 5-10\% higher than at the SPS).
This results in about 5\% flatter pion spectra while the effect on the
proton slope at small $p_\perp$ is much larger and at least 25\%. Again 
the flow fits describe all available spectra very well up to transverse 
momenta of about 2-3\,GeV (which covers more than 99\% of all hadrons), 
and only at higher $p_\perp$ one begins to see evidence
for the power-law tails expected from hard QCD processes. Also at RHIC, 
preliminary $\Omega$ and $\bar\Omega$ spectra \cite{VanBuren:2002sp} are
consistent with the $\Omega$ fully participating in the hadronic flow.
A comparison with hydrodynamic predictions published in 
\cite{Huovinen:2001cy} works well if the $\Omega$ is assumed to decouple 
at $T_{\rm f}\simeq 135-140$\,MeV, i.e. only slightly before the pions 
freeze out at $T_{\rm f}\eq125-130$\,MeV \cite{Heinz:2002un}, but not 
if one assumes decoupling already at hadronization, 
$T_{\rm f}{\eq}T_c\eq170$\,MeV,
where the hydrodynamic model has not yet developed enough transverse flow. 
  
The extracted flow velocities at or above half the speed of light 
demonstrate that the collision fireball indeed undergoes a violent explosion 
-- the ``Little Bang''. It appears to be impossible to obtain such large
flow velocities without assuming initial energy densities and pressures
well above the critical value for deconfinement. This is already true
at the SPS \cite{Heinz:2000ba} but, as I show next, much more convincingly 
so at RHIC. 

\begin{figure}[h]
\begin{minipage}[t]{108mm}
\includegraphics[width=108mm]{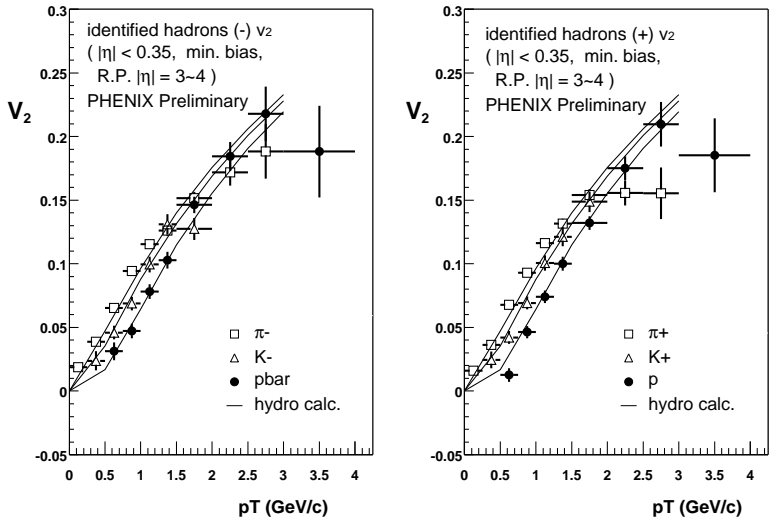}
\end{minipage}
\hspace{\fill}
\begin{minipage}[t]{48mm}
\vskip -77mm
\caption{Transverse momentum dependence of $v_2$ for identified 
particles, $\pi^-$, $K^-$, $\overline{p}$ (left) and $\pi^+$, 
$K^+$, $p$ (right), as measured by PHENIX \cite{Esumi:2002vy}. The 
solid lines represent the hydrodynamic prediction \cite{Huovinen:2001cy} 
for a freeze out temperature $T_{\rm f}\eq120$\,MeV 
for (from top to bottom) $\pi$, $K$, and $p$, respectively.}
\label{F1}
\end{minipage}
\end{figure}

A decisive measurement is the observation at RHIC, first made by STAR 
\cite{Ackermann:2001tr} and then confirmed by PHENIX 
\cite{PHENIXv2,Esumi:2002vy} and PHOBOS \cite{PHOBOSv2}, that the 
elliptic flow $v_2$ in non-central collisions is large and,
for transverse momenta below $p_\perp{\,\simeq\,}2$\,GeV, almost 
exhausts the upper limit provided by earlier hydrodynamic predictions
\cite{Kolb:2000sd,Huovinen:2001cy,PLB500} (see Fig.~\ref{F1}). The 
agreement between theory and data requires that the hydrodynamic 
evolution starts no later than about 1\,fm/$c$ after nuclear impact
\cite{Kolb:2000sd,Huovinen:2001cy} (the successful predictions
in \cite{Kolb:2000sd,PLB500} use $\tau_{\rm therm}\eq0.6$\,fm/$c$).
Since the spatial deformation responsible for the creation of flow 
anisotropies quickly decreases, $v_2$ develops and saturates early 
in the collision \cite{Sorge:1997pc,Kolb:2000sd}, long before hadrons 
decouple. Hence, the conclusion that the measured large $v_2$ implies 
early thermalization is not changed \cite{Teaney:2001av} if the somewhat 
unrealistic sharp ``Cooper-Frye freeze-out'' used in the hydrodynamic 
approach is replaced by a proper kinetic treatment of the freeze-out stage
\cite{Bass:2000ib}. The agreement of $v_2(p_\perp)$ out to transverse
momenta of about 2\,GeV, including the predicted \cite{Huovinen:2001cy} 
splitting of $v_2$ with the hadron rest mass (see Fig.~\ref{F1}), shows 
that the bulk of the matter (i.e. ${\,>\,}99{\%}$ of the emitted hadrons) 
behaves hydrodynamically. (One should note, however, that the measured 
rapidity dependence of 
\vskip -10mm
\vspace*{-8mm}
\noindent
$v_2$ \cite{PHOBOSv2} cannot be
reproduced by existing hydrodynamic models \cite{Hirano:2002ds}.) 
Microscopic models share this success only if they approach the 
hydrodynamic limit by invoking unusually strong rescattering among 
the fireball constituents \cite{Molnar:2001ux,Lin:2001zk,Humanic:2002iw}, 
far above the level conventionally expected from perturbative QCD. It 
is important to realize that, given the observed final multiplicity and 
transverse energy per unit rapidity, the maximum energy density in the 
fireball at $\tau_{\rm therm}\approx 0.6$\,fm/$c$ is about 25\,GeV/fm$^3$ in 
central collisions. Even after averaging over the transverse plane this 
is still more than an order of magnitude above the critical value for 
deconfinement and corresponds to about twice the critical temperature! 
Unless a completely different mechanism for creating the elliptic flow 
can be found, the conclusion seems unavoidable that a thermalized 
QGP at $\e{\,>\,}10\,\e_c$ and $T{\,\geq\,}2T_c$ is created at RHIC 
which lives for about 5-7\,fm/$c$ before becoming sufficiently dilute 
to form hadrons. Perturbative mechanisms \cite{Baier:2000sb} seem 
unable to explain the phenomenologically required very short 
thermalization time scale, pointing to strong non-perturbative 
dynamics in the QGP even at or above $2T_c$. 

\vspace*{-0.2cm}

\section{STATISTICAL HADRONIZATION: MEASURING $T_c$}
\label{sec2}

\vspace*{-0.1cm}

The quark-hadron phase transition is arguably the most strongly coupled
regime of QCD. Soft hadronization happens through a multitude of different
channels and is therefore most efficiently described in a statistical 
approach, as realized by Hagedorn more than 35 years ago \cite{Hagedorn:st}.
The microscopic processes are only constrained by local conservation
laws (valid inside causally connected volume regions $\Delta V$) for 
energy, baryon number and strangeness. Maximizing the entropy subject to 
these constraints results in local thermal and chemical ``equilibrium'' 
distributions for the hadrons \cite{Slotta:aw} whose local temperature 
$T_{\rm chem}$ and chemical potentials $\mu_B,\mu_S$ arise from Lagrange 
multipliers and reflect the local energy, baryon and strangeness densities 
at hadronization. This ``equilibrium'' is {\em not} achieved kinetically 
(by hadronic rescattering), but statistically (by interference of many 
channels). It does not require the prehadronic state to be thermalized, 
although {\em very} strong deviations from thermal equilibrium (e.g. at 
high $p_\perp$) may survive the hadronization process. The statistical 
approach is {\em only} expected to work for soft hadron production.

We know from lattice QCD that a hadronic equilibrium distribution at
$\e{\,>\,}\e_c$ is unstable against deconfinement, so statistical
hadronization can only proceed once the energy density has dropped to 
$\e_{\rm had}\eq\e_c{\,\approx\,}1$\,GeV/fm$^3$ \cite{Becattini:1997rv}.
The resulting equilibrium distribution will thus be characterized by 
$T_{\rm chem}{\eq}T_c$. Hadronic cascade models have shown 
\cite{Bass:1999tu} that, at small net baryon density, the total 
particle density below $T_c$ is too small and the expansion of the 
collision fireball is too rapid to maintain chemical equilibrium by 
inelastic hadronic collisions. Hence, the hadron abundances decouple 
directly at hadronization, and the chemical freeze-out temperature 
extracted from a statistical model fit to the measured abundances 
reflects directly the (de)confinement temperature, 
$T_{\rm chem}{\eq}T_c$ \cite{Tchem}. 

\vskip -7mm
\begin{figure}[h]
\begin{minipage}[t]{108mm}
\includegraphics[width=108mm,height=80mm]{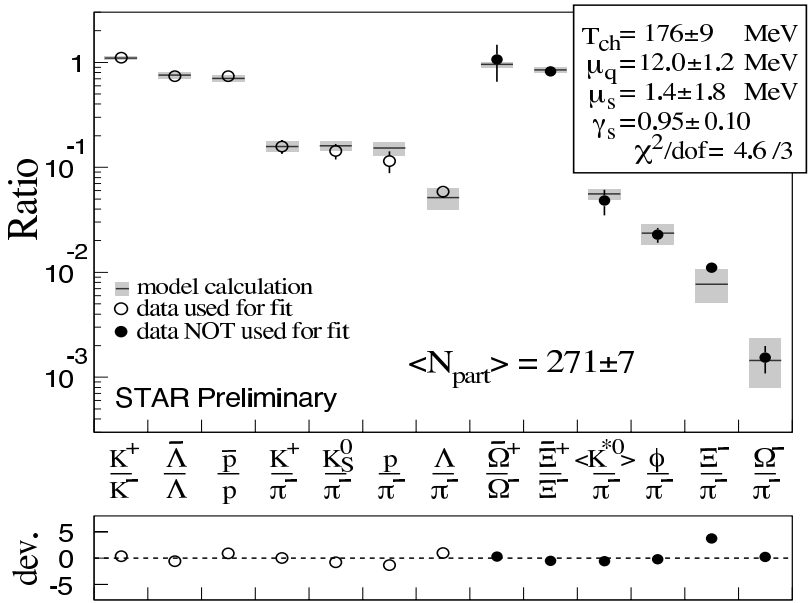}
\end{minipage}
\hspace{\fill}
\begin{minipage}[t]{45mm}
\vskip -80mm
\caption{Particle ratios from 130\,$A$\,GeV Au+Au collisions,
measured by the STAR Collaboration at RHIC \cite{VanBuren:2002sp} 
and fitted by M.~Kaneta to a thermal statistical model 
\cite{Letessier:1993hi,Braun-Munzinger:2001ip}. Hadron ratios measured 
by the BRAHMS and PHENIX Collabora\-tions agree with the shown STAR data.
}
\label{F2}
\end{minipage}
\end{figure}
\vspace*{-6mm}

Figure~\ref{F2} \cite{VanBuren:2002sp} shows that the RHIC data confirm 
this expectation, just as it was previously confirmed at the SPS 
\cite{Braun-Munzinger:1999qy}: the chemical decoupling 
temperature extracted from the particle ratios is consistent with the 
critical temperature for deconfinement, $T_c{\,\simeq\,}170$\,MeV, 
extracted from lattice QCD simulations \cite{Karsch_QM01}.
Fitting the maximum entropy parameters to some of the more abundant 
particle yields one is able to reproduce {\em all} measured particle
ratios, including those involving the rare multistrange (anti)baryons;
this is quite impressive. {\em Thermal freeze-out} of the momentum 
distributions is delayed by strong quasi-elastic scattering via hadron
resonances, such as $\pi+N\to\Delta\to\pi+N$, $\pi+N\to\Delta\to\pi+N$,
$\pi+K\to K^*\to\pi+K$, etc., which do not modify the total measured
yields of pions, kaons, and nucleons but continue to adjust their 
momentum distributions to the falling temperature and growing 
collective transverse flow until also resonance scattering ceases. 
The momentum spectra thus reflect a lower freeze-out temperature
$T_{\rm f}{\,<\,}T_{\rm chem}$, as seen in section~\ref{sec1}. Note that 
only the {\em total} abundances (after all decays of unstable resonances 
are taken into account, e.g. $N_\pi^{\rm tot}{\eq}N_\pi+2N_\rho +
 N_\Delta+N_{K^*}+\dots$) freeze out at $T_{\rm chem}$ while the 
fraction of pions stored in resonances still changes: due to their 
strong coupling to the cooling pion fluid, the resonance abundances keep
readjusting to the decreasing temperature, and the larger their pionic
decay width the later they decouple. This implies that the measured 
$K^*/K$, $\Delta/N$, $\Lambda^*/\Lambda$ etc. ratios should generally be 
smaller than their chemical equilibrium values at $T_{\rm chem}{\eq}T_c$ 
and, when themselves translated into a decoupling temperature, reflect 
more closely the spectral temperature $T_{\rm f}$ than $T_c$ 
\cite{Torrieri:2001ue}. First results on resonance/ground-state 
ratios were reported at {\em Quark Matter 2002} \cite{Fachini:2002yj}, 
and this prediction will be checked soon.

\vspace*{-0.2cm}

\section{JET QUENCHING: STRONG PARTON ENERGY LOSS IN THE QGP}
\label{sec3}

\vspace*{-0.1cm}

In 1982 Bjorken \cite{Bjorken:1982tu} suggested that fast partons 
travelling through a QGP might lose large amounts of energy by elastic 
scattering with the plasma constituents, resulting in the suppression of 
jets from the interior of the collision fireball in relativistic heavy ion 
collisions. Although the energy loss mechanism envisaged by him turned out
to be ineffective, his qualitative prediction appears to be impressively 
confirmed by recent RHIC data. A good review of the theory of parton energy 
loss was recently given by Baier \cite{Baier:2002tc}, and I refer to his 
talk for references. In a series of papers in the early 1990's, Gyulassy, 
Pl\"umer and X.N. Wang identified radiative energy loss as the dominant 
jet suppression mechanism, and in 1995-1997 Baier, Dokshitzer, Mueller,
Peign\'e and Schiff (BDMPS) realized that an important effect on the 
energy loss rate results from multiple color interactions of the 
radiated gluon with the colored plasma constituents \cite{Baier:1994bd}.
In the limit of an optically thick quark-gluon plasma they found 
\cite{Baier:1994bd} that the energy loss $\Delta E$ of the fast parton 
increases {\em quadratically} with the distance $L$ travelled before 
escaping and hadronizing,
\begin{eqnarray}
\label{1}
  \quad  
  \Delta E \approx \frac{\alpha_s}{2} \frac{\mu^2}{\lambda} \, L^2,
  \qquad{\rm with}\qquad 
  \frac{\mu^2}{\lambda} = \rho \int dq_\perp^2\, q_\perp^2\,
  \frac{d\sigma}{dq_\perp^2}\,,
\end{eqnarray}
where $\alpha_s$ is the strong coupling constant, $\rho$ is the plasma 
density, $\mu^2$ is the Debye screening mass for color electric fields 
in the plasma and $\lambda$ is the gluon mean free path. This was 
subsequently improved upon for plasmas with finite opacity by Gyulassy,
Levai, and Vitev \cite{Gyulassy:1999zd} and by Wiedemann 
\cite{Wiedemann:2000za}, using an opacity expansion. For 
optically thin plasmas this leads to an important dependence of
the energy loss $\Delta E$ on the energy $E$ of the fast parton
\cite{Gyulassy:1999zd} and to deviations from the quadratic path 
length dependence for small values of $L$. Both corrections reduce the
predicted energy loss relative to the BDPMS result (\ref{1}), in 
particular at ``low'' (SPS and RHIC) energies, but it remains still
significantly larger in a QGP than in subcritical normal hadronic 
matter \cite{Baier:2002tc}. With these improved results one may nurture 
the hope that parton energy loss and jet quenching, if observed, can be 
used as a quantitative probe of the density and its early time evolution 
in quark-gluon plasmas created in heavy-ion collisions.

Two important observations at RHIC have moved this hope to the brink of 
reality. The first is the discovery of a suppression of high-$p_\perp$
particle production in central Au+Au collisions at RHIC. When studied 
as a function of collision centrality, the production rates of charged 
hadrons at high $p_\perp{\,\gg\,}2$\,GeV were found 
\cite{Adcox:2001jp,Adler:2002xw,Mioduszewski:2002wt} {\em not} to follow 
the expected scaling with the number of binary collisions which 
characterizes hard QCD processes, but instead to continue to scale with 
$N_{\rm part}$ (the number of wounded nucleons), just as in the soft 
low-$p_\perp$ regime where phase cherence and the Landau-Pomeranchuk-Migdal
effect are known to suppress particle production. (There seems to be an 
intermediate $p_\perp$ region, 1.5\,GeV$<p_\perp{\,<\,}3.5$\,GeV, where 
particle production grows a bit faster than $N_{\rm part}$ but more slowly
than the number of binary collisions.) From the Glauber model the number
of binary $NN$ collisions is known to scale with $N_{\rm part}^{4/3}$ while
$N_{\rm part}$ is proportional to the volume of nuclear matter participating
in the collision. The observed suppression or ``quenching'' of hard particle
production by a factor $N_{\rm part}^{1/3}$ is thus proportional to the
linear size (``radius'') of the nuclear overlap volume. It is qualitatively
consistent with expectations from pQCD-based jet quenching calculations
\cite{Vitev:2002wh} and inconsistent with pQCD calculations without jet 
quenching. It is seen for charged and neutral pions 
\cite{Mioduszewski:2002wt} but apparently not for high-$p_\perp$ protons
\cite{DNP} (at least not in the presently accessible $p_\perp$ range for
proton PID); to what extent this ``lack of hard proton suppression'' is
simply a reflection of the observed increase of the $p/\pi$ ratio with
increasing $p_\perp$ \cite{Sakaguchi:2002bm} which, at least up to 
$p_\perp{\,\approx\,}2$\,GeV, is naturally explained by the stronger 
transverse flow effects on protons than on pions, is still unclear. 

\vskip -8.5mm
\begin{figure}[h]
\begin{minipage}[t]{108mm}
\includegraphics[width=108mm]{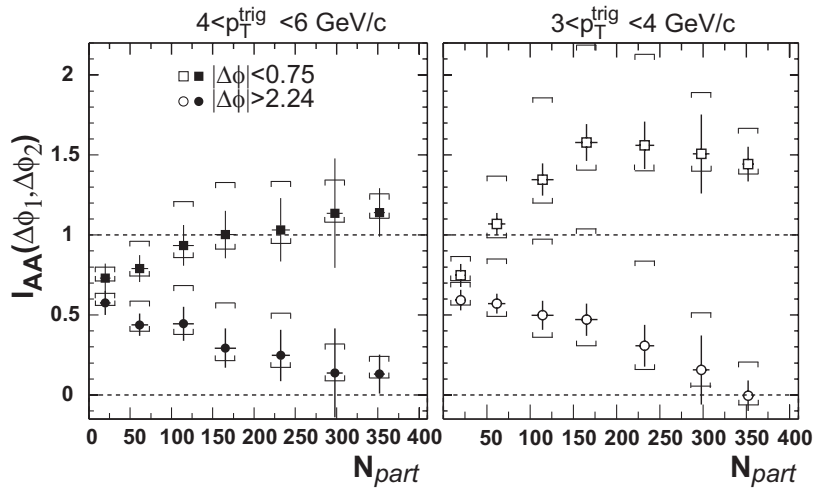}
\end{minipage}
\hspace{\fill}
\begin{minipage}[t]{50mm}
\vskip -72mm
\caption{Ratio of Au+Au and p+p of correlated pairs (see \cite{Adler:2002tq}
for details) for small (squares) and large (back-to-back, circles) 
azimuthal opening angles, for two different trigger $p_T$ intervals 
(left and right panels). The dominant systematic error 
arises from the uncertainty in the elliptic flow $v_2$ at these $p_T$.}
\label{F3}
\end{minipage}
\end{figure}
\vspace*{-9mm}

The second important observation, made by STAR \cite{Adler:2002tq} and 
(not yet as convincingly) by PHENIX \cite{Chiu:2002ma}, is that 
high-$p_\perp$ particle production is indeed due to jets, and that the 
observed overall suppression of high-$p_\perp$ particle production in
central Au+Au collisions goes hand-in-hand with the complete suppression
of the far-side partners of the observed jets. Perturbatively, hadron 
jets result from hard parton collisions leading to parton pairs with
large, approximately balancing transverse momenta which fragment into
pairs of back-to-back jets. In an angular distribution around the beam 
axis, jets appear in the two-particle correlation function as two 
peaks separated by 180$^\circ$. By triggering on a 
high-$p_\perp$ particle and correlating it with other particles of
$p_\perp{\,>\,}2$\,GeV (in order to suppress the uncorrelated soft 
background), these pairs of peaks were indeed identified in $pp$ and
peripheral Au+Au collisions at 200\,$A$\,GeV at RHIC \cite{Adler:2002tq},
but in {\em central} Au+Au collisions only the correlation peak at
small relative angles is found. Already in peripheral Au+Au collisions
the far-side peak at 180$^\circ$ is much smaller than expected from a 
superposition of $pp$ collisions \cite{Adler:2002tq,Chiu:2002ma}, and 
in central collisions it completely disappears after accounting for
the collective azimuthal correlations generated by elliptic flow
\cite{Adler:2002tq} (see Fig.~\ref{F3}).

While attempts to quantitatively understand these observations have only
just begun \cite{Muller:2002fa}, they suggest a very simple and intriguing
picture similar to the one already proposed in 1982 by Bjorken  
\cite{Bjorken:1982tu}: fast partons formed inside the hot and dense
collision zone suffer severe energy loss and become part of the
approximately thermalized, collectively expanding ``soup'' which 
emits particles at low $p_\perp{\,<\,}2$\,GeV. Only if the primary hard 
collision happens near the surface of the reaction zone, the ``outward 
moving'' parton fragments to become a jet, very much like in $pp$ collisions
(upper set of symbols in Fig.~\ref{F3}), whereas the ``inward moving'' 
parton is more and more efficiently absorbed by the medium as the
nuclei collide more centrally and the reaction fireball volume increases
(lower set of symbols in Fig.~\ref{F3}). For central Au+Au collisions and 
$p_\perp{\,<\,}8-10$\,GeV \cite{Adler:2002tq}, the ``inward moving'' 
parton doesn't make it to the other side with sufficient energy left to 
fragment into a recognizable jet. Jets are therefore only emitted from 
a relatively thin surface layer while the bulk of the fireball volume 
is opaque to jets. 

This very qualitative picture may appear naive but it nicely explains
as a surface/volume effect the missing factor $N_{\rm part}^{1/3}$ in the
observed scaling of high-$p_\perp$ particle production. It also allows
for a simple geometric estimate of the elliptic flow $v_2$ at high
$p_\perp$, resulting from the anisotropic emission of surface jets from 
the spatially deformed overlap region \cite{Shuryak:2001me}. As pointed 
out by Voloshin \cite{Voloshin:2002wa}, this simple geometric estimate 
\cite{Shuryak:2001me} appears to be consistent with the STAR data for 
$v_2$ at high $p_\perp$. Of course, at {\em very} high $p_\perp$ one
eventually expects the ``inward moving'' parton to emerge on the 
other side with sufficient energy left to form a jet. When this happens,
its energy loss can be studied as a function of its path length through
the medium, in order to check the energy loss formula (\ref{1}) and its
various published variants. Selecting events of fixed centrality (impact
parameter), the path length can be controlled geometrically by the azimuthal
angle between the jet and the reaction plane, and a new era of ``jet
tomography'' \cite{Gyulassy:2001nm} will begin.

\vspace*{-0.2cm}

\section{THE RHIC HBT PUZZLE}
\label{sec4}

\vspace*{-0.1cm}

Unfortunately, space-time limitations do not allow for a detailed discussion
of one piece of the puzzle which so far fails to seamlessly fit into the
picture painted above: the two-particle Hanbury Brown -- Twiss (HBT) 
correlation measurements. Since I reviewed these elsewhere (see
\cite{Heinz:2002un} and references therein), let me be brief here. HBT
correlations can be used to probe the size, shape and dynamical state of 
the source at hadronic decoupling. STAR and PHENIX have produced mutually
consistent results showing that the HBT radii in the sideward and outward 
directions (i.e. perpendicular and parallel to the transverse emission
vector) depend strongly on the transverse pair momentum and are almost equal
to each other. Otherwise successful dynamical models such as those discussed
in Sec.~\ref{sec1} fail to reproduce this strong transverse momentum
dependence and consistently give smaller sideward than outward radii, in 
contradiction with the data. They also overpredict the longitudinal radius.
The longitudinal and outward radii can be decreased by reducing the total
lifetime of the fireball, but this would require even faster thermalization
to accomodate the observed transverse flow, and the sideward radius remains 
too small. Equal outward and sideward radii not only require a very opaque
source (such as the hydrodynamic ones with unrealistic sharp Cooper-Frye
freeze-out), but also a short lifetime and a breaking of longitudinal
boost invariance, at least in the decoupling process. While these problems
are generally believed to reflect insufficiencies in the description
of the late decoupling stage which will not affect our understanding of 
the early collision (QGP) stage, we can't be sure of this until a model 
is found that works. 

\vspace*{-0.2cm}

\section{OUTLOOK}
\label{sec5}

\vspace*{-0.1cm}

The first two years of RHIC running have brought a rich harvest of 
hadron production data and, as I discussed, abundantly fulfilled our 
hopes and expectations of finding more and stronger evidence for the 
making of quark-gluon plasma in heavy-ion collisions. The observation 
of large elliptic flow, with its compelling interpretation in terms of
fast thermalization, and the discovery of jet quenching are two important
milestones which go significantly beyond what was earlier achieved at 
the lower SPS energy \cite{Heinz:2000bk} and which bring us closer to
a generally accepted ``proof'' of QGP creation. But much more is to come:
only now, with RHIC finally running at full energy and luminosity (and,
hopefully, for the full promised time per year) it is possible to address
such hallmark measurements as thermal dilepton and direct photon emission
and heavy quarkonium production, all of which play crucial roles in the 
early diagnostics of the QGP which we are apparently mass-producing at 
RHIC. While trying to solve the HBT puzzle and to quantitatively 
understand jet quenching, we are looking forward to these high-luminosity
measurements and any surprises they may bring.   



\end{document}